\documentclass{article}

\usepackage[a4paper]{geometry}
\usepackage{graphicx}
\usepackage[table]{xcolor}

\usepackage{amsmath}
\usepackage{amsfonts,amssymb}

\usepackage[T1]{fontenc}
\usepackage[latin1]{inputenc}
\usepackage{newtxtext,newtxmath,courier}
\usepackage{helvet}

\usepackage{caption}
\usepackage{cite}

\newcommand\authormark[1]{\textsuperscript{#1}}
\newcommand\address[1]{%
	{\raggedright\small\itshape #1\par}
}
\newcommand\email[1]{%
	{\raggedright\footnotesize\itshape#1\par}
}

\begin{document}

\title{Using fluorescent beads to emulate single flurophores}

\author{Luis A. Alem\'an-Casta\~neda,\authormark{1,*} Sherry Yi-Ting Feng,\authormark{2} Rodrigo Guti\'errez-Cuevas,\authormark{3}\\ Isael Herrera,\authormark{1} Thomas G. Brown,\authormark{2} Sophie Brasselet,\authormark{1} Miguel A. Alonso\authormark{1,2,**} }

\maketitle

\address{\authormark{1}Aix Marseille Univ, CNRS, Centrale Marseille, Institut Fresnel, Marseille 13013, France\\
	\authormark{2}The Institute of Optics, University of Rochester, Rochester, NY 14627, USA\\
	\authormark{3}Institut Langevin, ESPCI Paris, Universit\'e PSL, CNRS, 
	75005 Paris, France\\
}

\email{\authormark{*} aleman-castaneda@fresnel.fr, \authormark{**} miguel.alonso@fresnel.fr} 

\begin{abstract}
	We study the conditions under which fluorescent beads can be used to emulate single fluorescent molecules in the calibration of optical microscopes. Although beads are widely used due to their brightness and easy manipulation, there can be notable differences between the point spread functions (PSFs) they produce and those for single-molecule fluorophores, caused by their different emission pattern and their size. We study theoretically these differences for various scenarios, e.g. with or without polarization channel splitting, to determine the conditions under which the use of beads as a model for single molecules is valid. We also propose methods to model the blurring due to the size difference and compensate for it to produce PSFs that are more similar to those for single molecules.\\
\end{abstract}


\section{Introduction}
Fluorescence microscopy is a widely used technique for bio-imaging applications, not only due to its intrinsic optical contrast and/or specificity \cite{Mertz:2004} but also for its compatibility with super-resolution techniques\cite{Hell:1994, Rust:2006, Hess:2006}. For example, single-molecule localization microscopy (SMLM) can achieve a spatial resolution of up to a few nanometers \cite{Kao:1994, Hess:2006, Huang:2008, Liu:2013}. Additionally, when combined with polarization measurements, fluorescence microscopy can offer information about the sample's structural properties, in particular, the average orientation and the order/wobbling of the molecules, both on the ensemble and the single-molecule regimes \cite{Brasselet:2011}. This information is key not only for biological applications but also in a broader sense, since not considering the fluorophore's orientation can bias localization measurements\cite{Enderlein:2006, Backlund:2012}. The simultaneous estimation of the emitters' 3D localization and 3D orientational behavior is called single-molecule orientation localization microscopy (SMOLM). Common SMOLM  techniques include polarization channel splitting \cite{Stallinga:2012,ValadesCruz:2016, Caio:2022} and point spread function (PSF) engineering \cite{Backer:2014,Zhang:2018,Curcio:2020, Hullman:2021,Ding:2021, Wu:2022}.

Fluorescence from a single molecule can be modeled as dipolar radiation \cite{Bohmer:2003, Brasselet:2011}, resulting in a pattern at the pupil plane (or far field) that is not uniform in intensity nor in polarization, and it is strongly dependent on the dipole's 3D orientation \cite{Lieb:2004}. The resulting PSF at the detector plane can be predicted from this distribution. However, microscopes are typically imperfect optical systems that have optical aberrations and/or polarization distortions that bias both localization and orientation estimations. Furthermore, these distortions may change across the field of view. Therefore, it is important to characterize the imaging system properly, and one way to do so is by imaging fluorescent beads, given their brightness (significantly larger than that of single fluorophores) and easy manipulation \cite{Luchowski:2010, neumann:2013, zhang:2016,Petrov:2017,Li:2019,Yan:2019}. 

There are two main differences, however, between the PSFs for beads and those for actual single molecule emitters. First, while beads can be quite small (a few tens of nanometers), they are still much larger than single molecules and this spatial extension induces a blurring effect on the measured PSF. Second, beads are composed of many fluorophores with different orientations  \cite{Luchowski:2010,neumann:2013}, 
making their emission pattern quite different from that of a single dipole: the light emitted by a bead is in principle unpolarized and therefore fills fairly uniformly the pupil of the system, in contrast to single-molecule fluorophores that have specific polarization and intensity distributions.

In this paper, we study the circumstances under which fluorescent beads, with the help of polarizers or other optical components at the pupil plane, can be used to produce PSFs that mimic those for single fluorophores with different orientations. We start by discussing in Section~\ref{sec:fixed_dipole} the emission of a fixed dipole and in Section~\ref{sec:unpol} that of an unpolarized point-like emitter. The effect of bead size are then studied in Section~\ref{sec:beadsize}, both through numeric and semi-analytic approaches. We consider two simple models for the bead emission. In particular, the semi-analytic approach 
offers a series of corrections that can be used to counter the bead size effects on the PSF in image post-processing. Finally, the validity of the use of fluorescent beads supplemented with polarizers to mimic single fluorophores is discussed in Section~\ref{sec:mimicking}, and some conclusions are drawn in Section~\ref{sec:conclusions}. 
The analysis presented here follows directly from several theoretical advances due to Prof. Emil Wolf, such as the Richards-Wolf method for modeling high numerical aperture systems \cite{Richards:1959} and the theory for partial spatial coherence and polarization \cite{Born:coherence_book}, so we think this work is appropriate for this special issue celebrating his hundredth birthday.

\section{Radiation from a fixed dipole}\label{sec:fixed_dipole}
\begin{figure}[ht]
	\centering
	\includegraphics[scale=0.05]{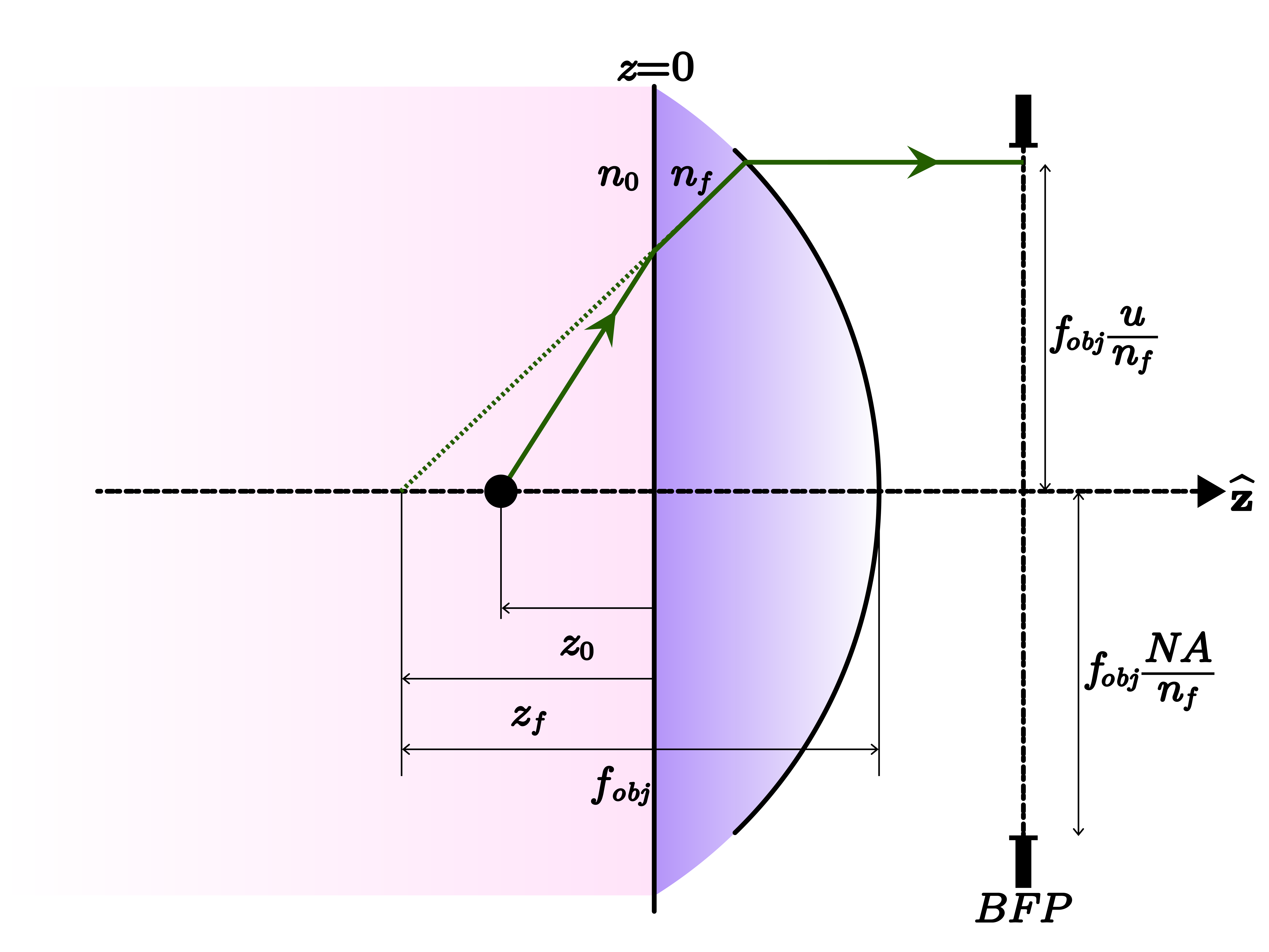}
	\caption{Sketch of the optical system collecting light from the fluorophore to the system's pupil. The fluorophore (black dot), located at an axial position $z_0$, is embedded in a medium with refractive index $n_0$. This medium has a planar interface at $z=0$ with a coverslip, whose refractive index $n_f$ is assumed to match that of the immersion liquid of the objective. The curved surface represents the collimation effect of the aplanatic objective, whose focal plane is at $z_f$. 
		The radial coordinate of a ray at the back focal plane (BFP), which coincides with the pupil plane, is given by $f_\mathrm{obj} ~u/n_f$.}
	\label{fig:1}   
\end{figure}

In this section we describe the forward model based on the Richards-Wolf formalism \cite{Richards:1959, Novotny:book, Curcio:2020} used to calculate the PSF of a single dipole at a distance $-z_0$ from the coverslip and for a system in which the location of the focal plane with respect the coverslip is given by $z_f$, as shown in Fig. \ref{fig:1}. We assume that the dipole's orientation is fixed and aligned with the unit vector $\boldsymbol{\hat{\mu}}^{\rm T}=(\mu_x,\mu_y,\mu_z)$. 
We ignore for now transverse ($x,y$) displacements, since they correspond simply to a linear phase factor at the pupil plane, and in turn to a transverse displacement at the detector plane. Let us denote as $\mathbf{E}_\mathrm{p}(\mathbf{u})$ the field at the system's pupil plane, where $\mathbf{u}$ is a dimensionless pupil position with polar coordinates $(u,\varphi)$, normalized so that the edge of the pupil corresponds to $u_\mathrm{max}=\mathrm{NA}$, the numerical aperture of the objective. For an ideal aplanatic system, the field at the pupil plane is given by 
\begin{equation}
	\mathbf{E}_\mathrm{p}(\mathbf{u};z_0,z_f)=e^{\mathrm{i}kn_fz_f\gamma(u,n_f)}e^{-\mathrm{i}kn_0z_0\gamma(u,n_0)}~\mathbb{K}(\mathbf{u})\cdot\boldsymbol{\hat{\mu}},
\end{equation}
where $n_0$ is the refractive index of the medium embedding the fluorophore, $n_f$ is that of the immersion liquid of the objective (assumed here to be the same as that of the coverslip), and
$\gamma(u,n)=\sqrt{1-(u/n)^2}$ \cite{Richards:1959}. Here, $\mathbb{K}(\mathbf{u})$ is a $2\times3$ matrix that determines the amplitude/polarization distribution at the pupil plane (where the field is paraxial and hence requires only two components) generated by each of the three Cartesian components of the dipole source, according to
\begin{align}\label{eq:Kmatrix}
	\mathbb{K}(\mathbf{u})&=\begin{pmatrix}
		g_0(u)+g_2(u) \cos2\varphi & g_2(u) \sin2\varphi & g_1(u)\cos\varphi\\
		g_2(u)\sin2\varphi &g_0(u)-g_2(u)\cos2\varphi & g_1(u)\sin\varphi
	\end{pmatrix},
\end{align}
where
\begin{subequations}
	\begin{align}
		g_{0,2}(u)&=P(u;n_f,n_0)\left[t_p(u)\gamma(u,n_0)\pm t_s(u)\right],\\
		g_1(u)&=2 P(u;n_f,n_0)~\frac{u}{n_0}~t_p(u),\\
		P(u;n_f,n_0)&=\frac{1}{2}\frac{n_f}{n_0}\frac{\sqrt{\gamma(u,n_f)}}{\gamma(u,n_0)},
	\end{align}
\end{subequations}
with $t_p(u)$ and $t_s(u)$ being the transmission coefficients for the radial (transverse-magnetic) and azimuthal (transverse-electric) components of the field, respectively.

\begin{figure}[ht]
	\centering
	\includegraphics[scale=0.8]{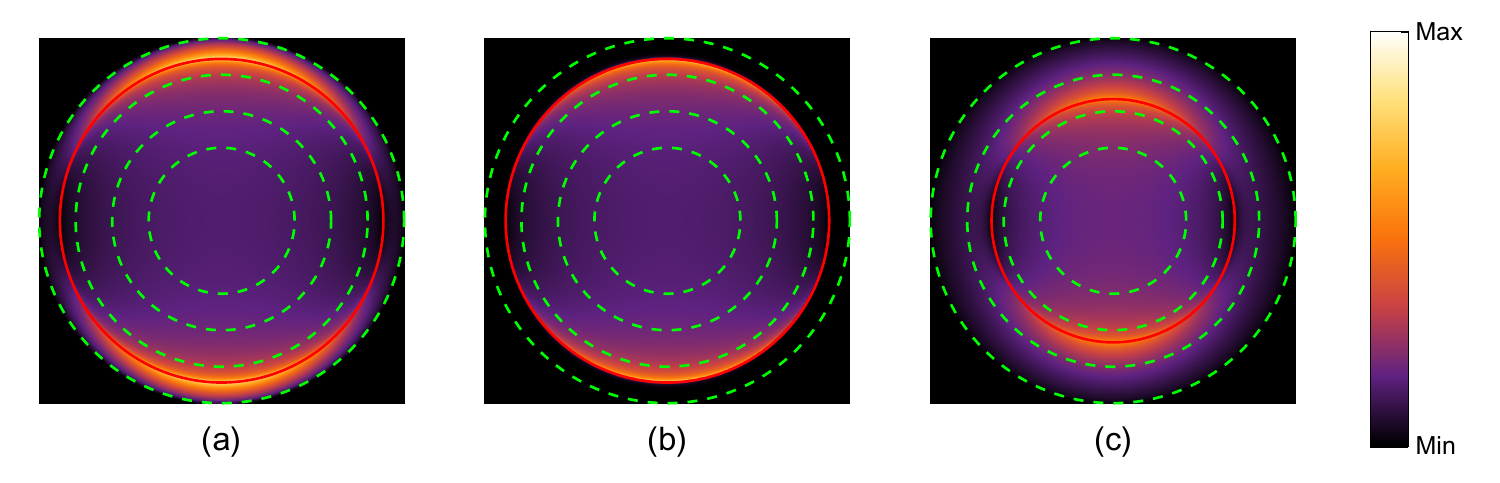}
	\caption{Intensity distribution at the pupil 
		generated by a horizontal dipole embedded in (a,b) water ($n_0=1.33$), and (c) air ($n_0=1$). For all $n_f=1.51$. The dashed green circles denote (from smaller to larger) $\mathrm{NA}=0.6,~0.9,~1.2$ and $1.5$, while the red circle indicates $\mathrm{NA}=n_0$, which is the radial lower bound of the annular region corresponding to SAF. In (a) and (c) SAF is important since the dipole is close to the coverslip, $|z_0|=\lambda/10$, while in (b) SAF is negligible since the dipole is further away, $|z_0|=\lambda$. 
	}
	\label{fig:2}   
\end{figure}

\begin{figure}[ht]
	\centering
	\includegraphics[scale=0.8]{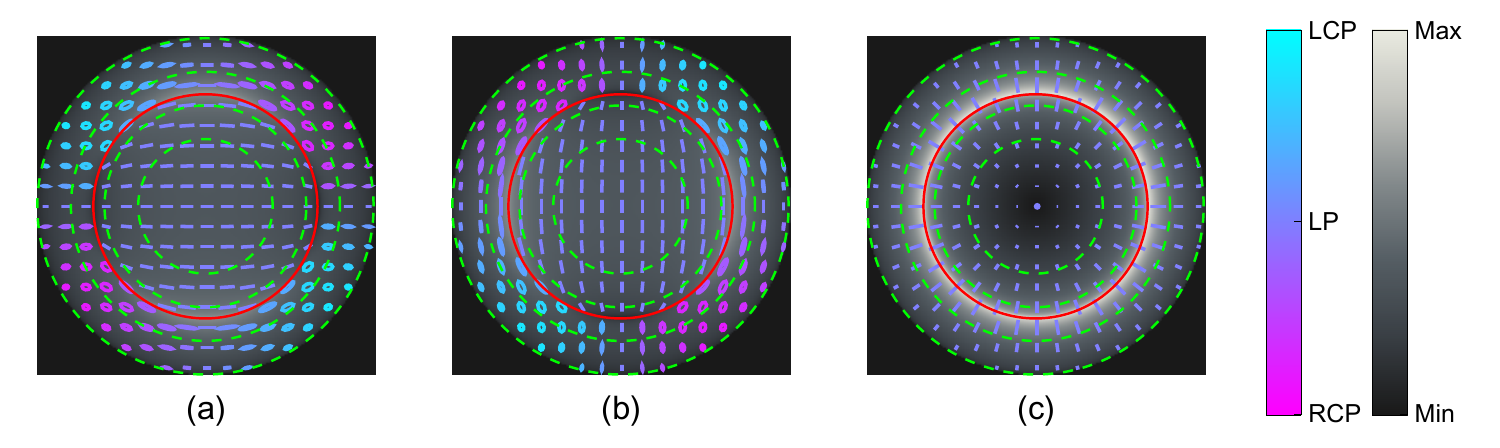}
	\caption{Polarization distribution at the pupil generated by a (a) transverse horizontal, (b) transverse vertical, and (c) longitudinal dipole embedded in air and with $n_f=1.51$.  Polarization ellipses are plotted on top of the respective amplitude distributions, with cyan denoting left-handedness and magenta right-handedness. The dashed green circles denote (from smaller to larger) $\mathrm{NA}=0.6,~0.9,~1.2$ and $1.5$, while the red circle indicates $\mathrm{NA}=n_0$, where SAF begins.  Intensity distributions are normalized with their respective maxima.
	}
	\label{fig:3}   
\end{figure}

Figure \ref{fig:2} shows the intensity distribution for a horizontal dipole embedded either in water (a,b) or in air (c), and at different distances from the coverslip. Notice that we observe supercritical angle fluorescence (SAF) if the NA of the system is larger than the refractive index of the embedding medium. The strength of SAF depends greatly on the distance between the dipole and the coverslip, $|z_0|$, being significant only for $|z_0|<\lambda$. Figure \ref{fig:3} shows the polarization distribution at the pupil for three orthogonal dipole orientations in air. Polarization within the pupil is everywhere linearly polarized except in the annular region occupied by SAF (outside the red circle), in which $t_p$ and $t_s$ become complex and introduce a relative phase between the radial and azimuthal field components. For a longitudinal dipole, however, polarization remains linear even in the SAF region since only the radial component contributes. Note that for transverse dipoles, polarization at the pupil is largely in the dipole direction for low NA, but less so for higher NA. 

The field from the pupil is focused on the detector, which in the ideal case corresponds to a Fourier transform of the form
\begin{equation}
	\mathbf{E}_\mathrm{det}(\boldsymbol{\bar{\rho}};z_0,z_f)=\frac{f}{M^2 \lambda }\int\mathbb{J}(\mathbf{u})\cdot\mathbf{E}_\mathrm{p}(\mathbf{u};z_0,z_f)\exp\left(-\mathrm{i}k\mathbf{u}\cdot\frac{\boldsymbol{\bar{\rho}}}{M}\right)d^2\mathbf{u},
\end{equation}
where $M$ is the optical system's magnification, $f$ is the focal length of the focusing system, and $\boldsymbol{\bar{\rho}}$ is the transverse position variable at the detector. We employ barred variables to distinguish the detector coordinates from the object space of the fluorophores. Here, $\mathbb{J}(\mathbf{u})$ represents a Jones matrix describing any manipulation on the field performed after the pupil plane, ranging from a simple uniform analyzer before the detector to a more elaborate mask with varying amplitude, phase or birefringence, like those used in many PSF engineering techniques \cite{Petrov:2017,Li:2019,Yan:2019,Curcio:2020, Ding:2021, Wu:2022}.  In what follows we use the shorthand $\mathcal{F}$ for this Fourier transform including the corresponding physical parameters. The final intensity at the detector is given by
\begin{equation}
	I_\mathrm{det}(\boldsymbol{\bar{\rho}};z_0,z_f)=\left[\mathcal{F}\left\{\mathbb{J}(\mathbf{u})\cdot\mathbf{E}_\mathrm{p}(\mathbf{u};z_0,z_f)\right\}\right]^\dagger\left[\mathcal{F}\left\{\mathbb{J}(\mathbf{u})\cdot\mathbf{E}_\mathrm{p}(\mathbf{u};z_0,z_f)\right\}\right],
\end{equation}
where $^\dagger$ denotes the conjugate transpose operation.

We now study the variation of the PSF with the axial position $z_0$ of the dipole. For this purpose we
expand in a Taylor series in $z_0$ the intensity at the detector plane. 
This expansion will let us 
gain physical insight and compare
different sources such as fixed dipoles, unpolarized point-like emitters, and fluorescent beads. It will also be the basis for a semi-analytic model for accounting for bead size discussed in Section \ref{sec:beadsize}. Let this Taylor expansion be centered around some value $z_0=d$: 
\begin{equation}
	\label{eq:TaylorSeries}
	I_\mathrm{det}(\boldsymbol{\bar{\rho}};z_0,z_f)=\sum_{m=0}^{\infty}\frac{1}{m!}(z_0-d)^m\frac{\partial^m I_\mathrm{det}}{\partial z_0^m}(\boldsymbol{\bar{\rho}};d,z_f)=\sum_{m=0}^{\infty}\frac{1}{m!}(z_0-d)^m\mathcal{I}_m(\boldsymbol{\bar{\rho}};d,z_f),
\end{equation}
where in the last step we introduced the notation $\mathcal{I}_m(\boldsymbol{\bar{\rho}};d,z_m)$ for the $m$th derivative in $z_0$ of the intensity. While written here as a sum over all non-negative integers, in practice the sum in $m$ will be truncated at some modest value.  Since the only dependence in $z_0$ of the field $\mathbf{E}_\mathrm{det}$ is within the complex phase factor, we have
\begin{equation}
	\frac{\partial^m\mathbf{E}_\mathrm{det}}{\partial z_0^m}(\boldsymbol{\bar{\rho}};d,z_f)=\mathcal{F}\left\{\mathbb{J}(\mathbf{u})\cdot\mathbf{E}_\mathrm{p}(\mathbf{u};d,z_f) [-\mathrm{i}kn_0\gamma(u,n_0)]^m\right\},
\end{equation}
so that the coefficients of the series 
become 
\begin{equation}
	\begin{aligned}
		\mathcal{I}_m(\boldsymbol{\bar{\rho}};d,z_f)=\sum_{l=0}^m\frac{m!}{l!~(m-l)!}\frac{\partial^l\mathbf{E}_\mathrm{det}^\dagger}{\partial z_0^l}(\boldsymbol{\bar{\rho}};d,z_f)\frac{\partial^{m-l}\mathbf{E}_\mathrm{det}}{\partial z_0^{m-l}}(\boldsymbol{\bar{\rho}};d,z_f).
	\end{aligned}
\end{equation}

\begin{figure}[ht]
	\centering
	\includegraphics[scale=1.4]{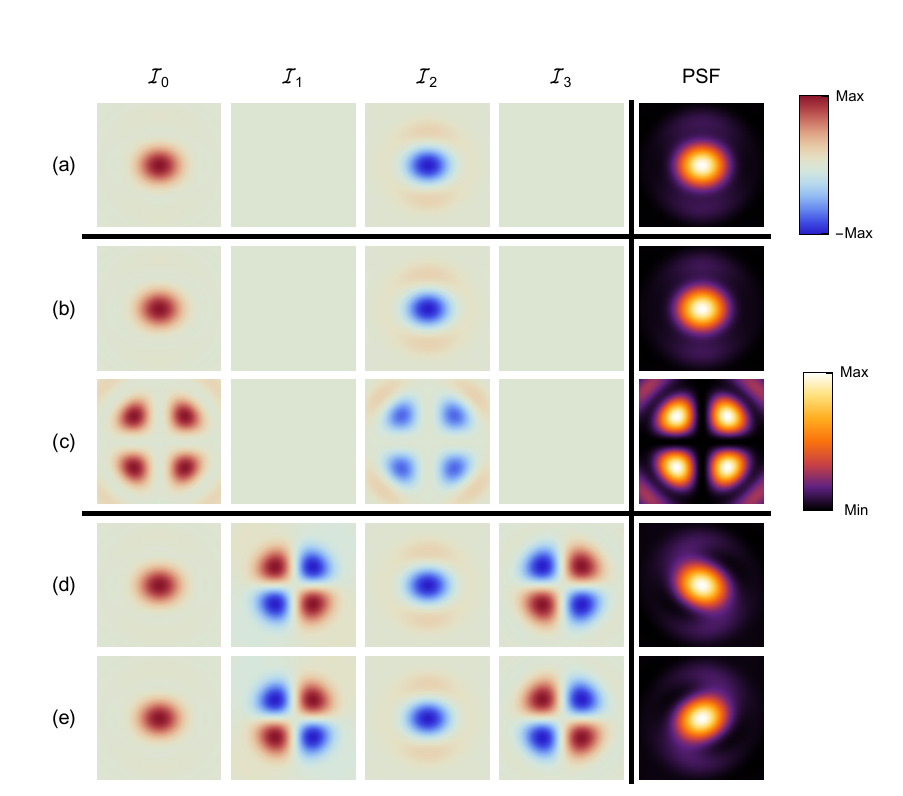}
	\caption{First four columns are the PSFs ${\cal I}_0$ and the subsequent Taylor components ${\cal I}_m$ for a system with NA$=1.3$ (no SAF) imaging a horizontal dipole in water in an index-matching scenario ($n_0=n_f\approx1.33$), for several polarization projections following the pupil: (a) no polarizer, (b) horizontal, (c) vertical, (d) left-circular and (e) right-circular polarizers. The last column shows the corresponding simulated defocused PSFs at $
		z_f-z_0=\lambda/3$. The elements of each column are normalized to the largest value in the column, but the results in (c) are scaled by a factor of 128 and those in (d) and (e) are scaled by a factor of 2.  Here $\lambda=525$nm and squares are of the length of $1\mu$m on the object side.} 
	\label{fig:4}   
\end{figure}

Note that for $m>0$, $\mathcal{I}_m$ are not intensities but rather changes in the intensity with $z_0$, so they can be positive or negative, although their individual integrals must vanish due to energy conservation. To illustrate the usefulness of this expansion, let us consider the simple case of index-matching (a common strategy in microscopy
to avoid optical aberrations introduced by the interface), i.e. $n_0=n_f$, for which there is no SAF. The first four columns of Fig.~\ref{fig:4} show the distributions for $\mathcal{I}_m$ at the nominal focal distance ($z_0=z_f$) for different constant choices for the Jones matrix $\mathbb{J}$, namely the identity (no manipulation before focusing), and pairs of orthogonal polarization projections before the detector: horizontal/vertical and circular left/right. The fifth column shows the PSFs at the defocused plane $z_f-z_0=\lambda/3$, calculated as a weighted superposition of the first four columns following Eq.~(\ref{eq:TaylorSeries}). 
Notice that the PSF is slightly elongated in the dipole direction (horizontally). 
Given the absence of SAF contributions and aberrations, the focal spot is symmetric in $z$ around the nominal image plane, so that the odd $m$ contributions vanish in Fig.~\ref{fig:4}(a-c). 
Nevertheless, as shown in Fig.~\ref{fig:4}(d,e) when a circular analyzer is used, ${\cal I}_1$ does not vanish, hence breaking the symmetry, 
and causing 
the circularly-polarized PSFs to counter-rotate with respect to each other with defocus, as shown in the last column of Fig.~\ref{fig:4}(d,e). This could serve as the basis of a SMOLM technique for the estimation of transverse dipoles' localization and orientation, consisting of forming separately the PSF of each circular component: for perfectly in-focus dipoles the elongation reports on the dipole direction, while for defocused dipoles the amount of counter-rotation between the PSFs reports on the defocus. This behavior is reminiscent of that of the PSFs in a technique proposed recently \cite{Curcio:2020}. However, the signature of the elongation and rotation of the PSFs in Fig.~\ref{fig:4}(d,e) may not be sufficiently strong to remain detectable in the presence of noise and camera pixelation. 

This index-matching example shows some of the insights gained by this Taylor expansion approach. In particular, it illustrates that in order to break the insensitivity to the sign of defocus of the measured PSFs, it is not imperative to have optical aberrations or use complicated Fourier manipulations (a common belief), but one can simply use the correct polarization projections. Note that the distributions for $\mathcal{I}_m$ when the refractive indices are not matched ($n_0\neq n_f$) can also be computed, even when SAF contributions are present.
These contributions would disrupt the symmetry around the focal plane so that $\mathcal{I}_1$ would not vanish. In fact, the choice of a focal plane would no longer be unique, and some criterion (e.g.~paraxial focus, minimmun spot size, minimum rms wavefront error, etc.) would have to be adopted for its choice. Nonetheless, the insight gained in the index-matching case remains valid, and the PSFs resulting from circular polarization projection would still counter-rotating with defocus. 

\section{Radiation from an unpolarized emitter.}
\label{sec:unpol}

\begin{figure}[ht]
	\centering
	\includegraphics[scale=0.8]{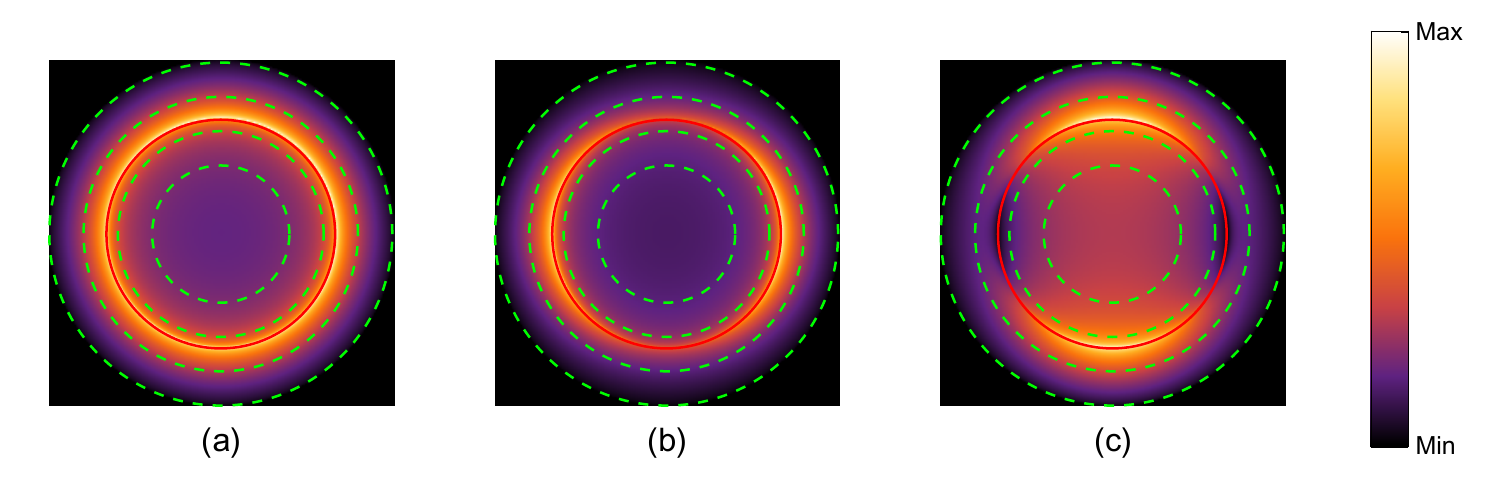}
	\caption{Intensity distributions at the pupil for (a) an unpolarized emitter, (b) an unpolarized emitter filtered by a horizontal polarizer at the pupil, and (c) a horizontal dipole, all embedded in air, with a distance to the coverslip of $d=\lambda/10$, and for a system with NA$=1.5$. The dashed green circles denote (from smaller to larger) $\mathrm{NA}=0.6,~0.9,~1.2$ and $1.5$, while the red circle indicates ${\rm NA}=n_0$ where SAF begins. }
	\label{fig:5}   
\end{figure}

An unpolarized (in the 3D sense) point-like emitter can be understood as the incoherent sum of the contributions from three orthogonal dipoles of equal magnitude. Without loss of generality, we can use three dipoles aligned with the coordinate axes $x$ (transverse horizontal), $y$ (transverse vertical), and $z$ (longitudinal). Therefore, the intensity distribution at the pupil can be written as
\begin{equation}
	I_\mathrm{pupil,unp}=|\mathbb{K}\cdot\mathbf{\hat{x}}|^2+|\mathbb{K}\cdot\mathbf{\hat{y}}|^2+|\mathbb{K}\cdot\mathbf{\hat{z}}|^2=\mathrm{Tr}\left\{\mathbb{K}^\dagger \mathbb{K}\right\},
\end{equation}
where $\mathbf{\hat{x}}$, $\mathbf{\hat{y}}$ and $\mathbf{\hat{z}}$ are 3D unit vectors in these three directions, and $\mathrm{Tr}\{\}$ denotes the trace. It can be shown that this intensity distribution reduces to the radially symmetric function
\begin{equation}
	I_\mathrm{pupil,unp}(u)=|P(u;n_f,n_0)|^2F(u;z_0)\left\{\mathcal{S}|t_p(u)|^2+|t_s(u)|^2\right\},
\end{equation}
where $F(u,z_0)=\mathrm{exp}[-2~ \mathrm{Im}\,\{kn_0|z_0|\gamma(u,n_0)\}]$  and $\mathcal{S}=|\gamma(u,n_0)|^2+\left(u/n_0\right)^2$, with $z_0$ the distance to the coverslip and $\mathrm{Im}\,\{\}$ denoting the imaginary part. Note that outside of the SAF region, i.e. for $u\leq n_0$, $P(u;n_f,n_0)=F(u,z_0)=1$. Figure \ref{fig:5} shows this intensity distribution and the one resulting from passing through a horizontal polarizer, as well as the one for a horizontal dipole. Note that the latter two are similar (i.e. fairly constant) for low NA but their differences increase considerably for higher NA. In particular, the distribution in Fig.~\ref{fig:5}(b) is maximal at the two intersections of the horizontal with the circle of radius $n_0$, and these points corresponds to the minima in Fig.~\ref{fig:5}(c).

\begin{figure}[ht]
	\centering
	\includegraphics[scale=0.8]{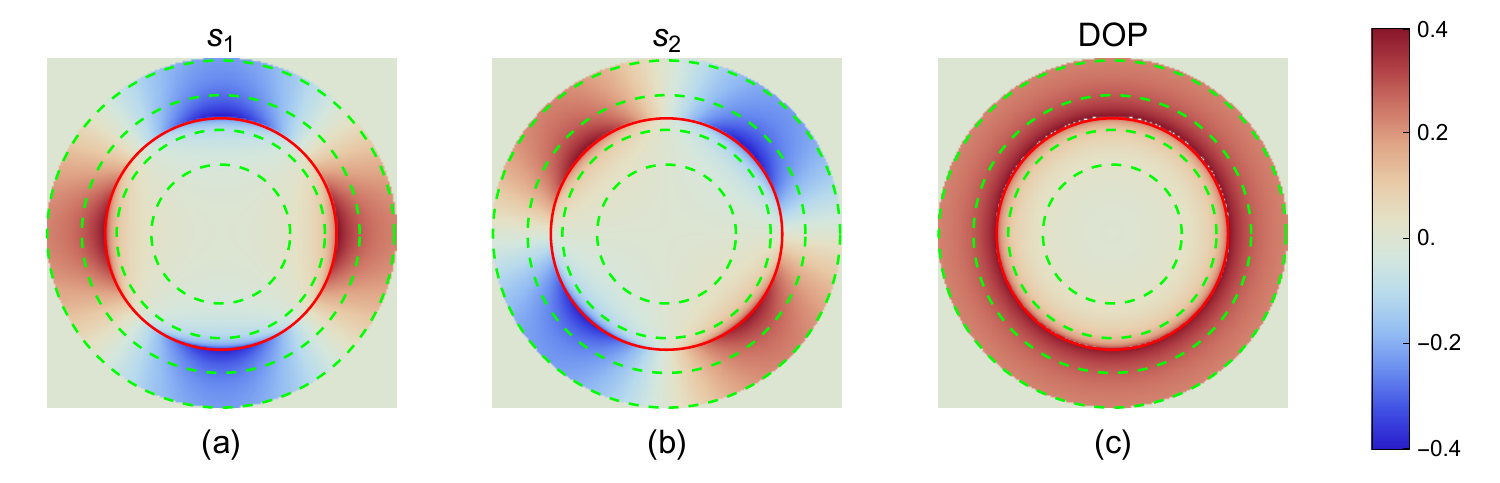}
	\caption{Normalized Stokes parameters at the pupil (a) $s_{1}$ and (b) $s_{2}$, and (c) $\mathrm{DOP_{\mathrm{p}}}$, over the pupil for an unpolarized emitter embedded in air, with a distance to the coverslip of $d=\lambda/10$, and for a system with NA$=1.5$. The dashed green circles denote (from smaller to larger) $\mathrm{NA}=0.6,~0.9,~1.2$ and $1.5$, while the red circle indicates ${\rm NA}=n_0$ where SAF begins. }
	\label{fig:6}   
\end{figure}

Let us now look at the polarization distribution at the pupil plane for an unpolarized point emitter. The distribution of the normalized Stokes parameters, i.e. $s_{i}(\mathbf{u})=S_{i}(\mathbf{u})/S_{0}(\mathbf{u})$ for $i=1,2,3$ where $S_{0}(\mathbf{u})=I_\mathrm{pupil,unp}(\mathbf{u})$,  are found to be given by
\begin{subequations}
	\begin{align}
		s_{1}(\mathbf{u})&=\frac{\mathcal{S}|t_p(u)|^2-|t_s(u)|^2}{\mathcal{S}|t_p(u)|^2+|t_s(u)|^2}\cos2\varphi,\\ 
		s_{2}(\mathbf{u})&=\frac{\mathcal{S}|t_p(u)|^2-|t_s(u)|^2}{\mathcal{S}|t_p(u)|^2+|t_s(u)|^2}\sin2\varphi,\\
		s_{3}(\mathbf{u})&=0,
	\end{align}
\end{subequations}
which (unlike $I_\mathrm{pupil,unp}$) depend on $\varphi$ although as a simple sinusoidal modulation. The degree of polarization is rotationally symmetric:
\begin{equation}
	\mathrm{DOP}(u)=\sqrt{\sum_{i=1}^3s_{i}^2(\mathbf{u})}=\frac{\left|\mathcal{S}|t_p(u)|^2-|t_s(u)|^2\right|}{\mathcal{S}|t_p(u)|^2+|t_s(u)|^2}.
\end{equation}
That is, the field at the pupil plane is fully unpolarized  ($\mathrm{DOP}=0$) only at the center, and as one approaches the border of the pupil the difference between the Fresnel transmission coefficients introduces more polarization. This degree of polarization can be particularly strong for the SAF contribution. This is illustrated in Fig. \ref{fig:6}, which shows the pupil distributions of the normalized Stokes parameters as well as the degree of polarization of an unpolarized emitter close to the coverslip. 

\begin{figure}[ht]
	\centering
	\includegraphics[scale=1.4]{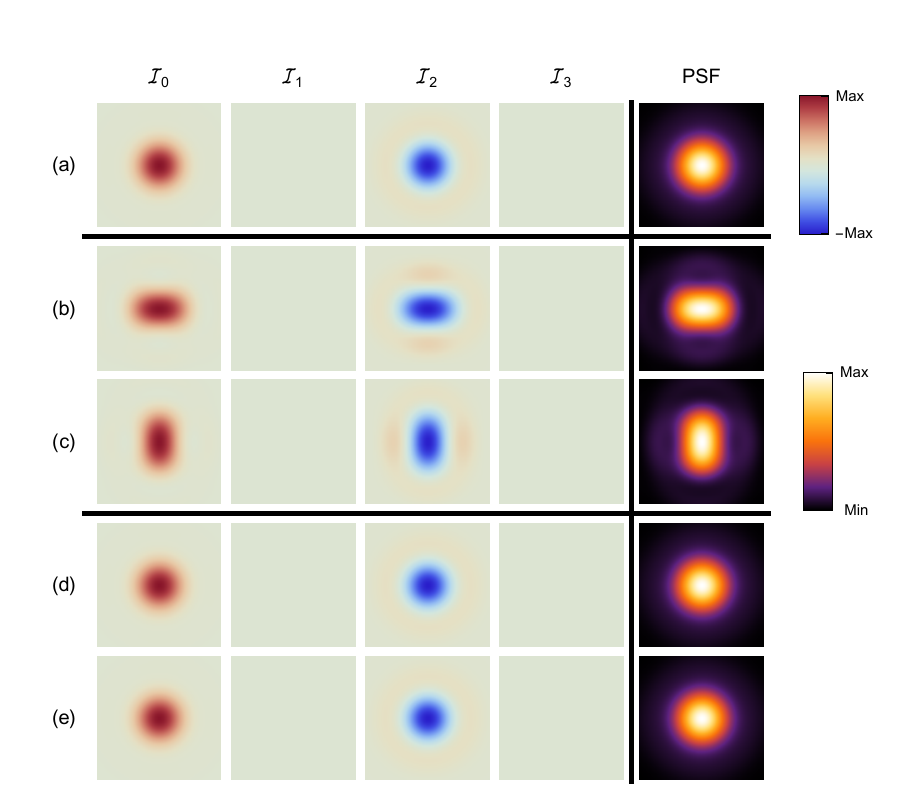}
	\caption{First four columns are the PSFs ${\cal I}_0$ and the subsequent Taylor components ${\cal I}_m$ for a system with NA$=1.3$ imaging an unpolarized emitter in water in a index-matching scenario ($n_f=n_0\approx1.33$), for several polarization projections following the pupil: (a) no polarizer, (b) horizontal, (c) vertical, (d) left-circular and (e) right-circular polarizers. The last column shows the corresponding simulated defocused PSFs at $z_f-z_0=\lambda/3$. The elements of each column are normalized to the largest value in the column, but a factor of 2 is applied in rows (b-e) with respect to (a). Here $\lambda=525$nm and squares are of length of $1\mu$m on the object side.}
	\label{fig:7}   
\end{figure}

Like the pupil distributions, the PSFs can be constructed as the incoherent sum of the intensity contributions at the detector of the three orthogonal dipoles (horizontal, vertical, and longitudinal). Note that, given the linearity of the intensity, the Taylor expansion for the total intensity is simply the sum of the corresponding contributions $\mathcal{I}_m(\boldsymbol{\bar{\rho}};z_0)$ for each of these three orthogonal dipoles. 
Figure \ref{fig:7} shows the distribution of $\mathcal{I}_m$ at the nominal focal distance, $z_0=z_f$, in an index-matching scenario, $n_0=n_f$, for different constant choices for $\mathbb{J}$, namely the identity (no manipulation before focusing), and pairs of orthogonal polarization projections before the detector: horizontal/vertical and left/right circular. For horizontal and vertical polarization projections, the PSFs become elongated, whereas they retain rotational symmetry when no polarization projection takes place or for circular polarization projections. Note that, given the symmetry of the focused field around the nominal image plane in the absence of aberrations, the odd derivative terms vanish in Fig.~\ref{fig:7}.


\section{Modeling the effects of bead size} \label{sec:beadsize}
As mentioned in the introduction, we can envision several simple models for the emission and geometry of a fluorescent bead, which include: (1) a \textit{transparent solid sphere}, in which all points inside the bead emit light, and this light is able to emerge from the bead; (2) a \textit{transparent shell}, in which only the points at the surface of the sphere emit, but the light from all these points is visible in any direction; (3) an \textit{opaque shell}, where only the light from points at the surface in line of sight can be detected, perhaps following a Lambertian emission model. Given their transparency, fluorescent beads most likely are best described by the first model, particularly when their index is well matched to that of its surrounding medium. However, we also describe the second model in what follows, given how similar its mathematical description is to that of the first. The third model will not be discussed further, since it is harder to model and would be relevant, for example, if the refractive index of the bead had an important imaginary part. 

To model an extended incoherent source we could follow several approaches. One would be to superpose incoherently the contributions from randomly oriented dipoles whose positions within the desired volume are also chosen randomly. This stochastic approach can be easily implemented for simulating fluorophore distributions of arbitrary shapes. However, it is computationally slow, and it does not offer much physical insight. We therefore opt instead for a deterministic approach that assumes the same emission at all points, hence allowing the use of numerical or semi-analytic approximations.

\subsection{Transparent solid sphere}
Let us assume that the intensity distribution at the detector due to a source at $(x,y,z)$ is $I_\mathrm{det}(\boldsymbol{\bar{\rho}};x,y,z)$ (where again coordinates without a bar describe the space of the bead, and barred coordinates correspond to the detector plane). Note that we omit the dependence on $z_f$ for brevity. Consider a transparent fluorescent bead of radius $R$ centered at $(x_0,y_0,z_0)$. The measured intensity can be calculated as a superposition of 2D convolutions:
\begin{equation}
	I_{\mathrm{tot}}(\boldsymbol{\bar{\rho}};x_0,y_0,z_0,R)=\sigma_V\int^{R}_{-R} I_{\mathrm{det}}(\boldsymbol{\bar{\rho}};x_0,y_0,z_0+z)\ast D(\boldsymbol{\bar{\rho}};R,z) \,dz,
	\label{eq:full_sphere}
\end{equation}
where $\sigma_V$ is the volumetric density of fluorophores, $\ast$ denotes a 2D convolution in the detector coordinates, and $D(\boldsymbol{\bar{\rho}};R,z)$ is a disk function representing the magnified section of the bead at $z$, where the transverse magnification is given by $M$, so
\begin{equation}
	D(\boldsymbol{\bar{\rho}};R,z)=\mathrm{circ}\left(\frac{1}{R}\sqrt{\left|\frac{\boldsymbol{\bar{\rho}}}{M}\right|^2+z^2}~\right)=\mathrm{circ}\left(\frac{|\boldsymbol{\bar{\rho}}/M|}{\sqrt{R^2-z^2}}\right).
\end{equation}
For numerical purposes, the integral in $z$ can be approximated through a discrete sum of $N$ slices, i.e.
\begin{equation}
	I_{\mathrm{tot}}(\boldsymbol{\bar{\rho}};x_0,y_0,z_0,R)\approx\sigma_V\sum_{j=1}^{N} I_{\mathrm{det}}(\boldsymbol{\bar{\rho}};x_0,y_0,z_0+z_j)\ast D(\boldsymbol{\bar{\rho}};R,z_j) \,\Delta z,
	\label{eq:sum_disks}
\end{equation}
where
\begin{equation}\label{eq:deltaz}
	\begin{aligned}
		\Delta z&=\frac{2R}{N}, & z_j&=\left(j-\frac{1}{2}\right)\Delta z-R.
	\end{aligned}
\end{equation}
This approximate model can then be implemented as
\begin{equation} \label{eq:slices}
	I_{\mathrm{tot}}=\sigma_V~\mathcal{F}^{-1}_{\mathbf{u}\rightarrow\boldsymbol{\bar{\rho}}}\left\{\sum_{j=1}^{N}\mathcal{F}_{\boldsymbol{\bar{\rho}}\rightarrow\mathbf{u}}\left\{ I_{\mathrm{det}}(\boldsymbol{\bar{\rho}};x_0,y_0,z_0+z_j)\right\}\mathcal{F}_{\boldsymbol{\bar{\rho}}\rightarrow\mathbf{u}}\left\{ D(\boldsymbol{\bar{\rho}};R,z_j) \right\} \right\}\Delta z,
\end{equation}
where
\begin{equation}
	\mathcal{F}_{\boldsymbol{\bar{\rho}}\rightarrow\mathbf{u}}\left\{ D(\boldsymbol{\bar{\rho}};R,z_j) \right\}(u)=2\pi M^2 R_j^2 ~ \frac{J_1(kuR_j)}{kuR_j},
\end{equation}
with $R_j=\sqrt{R^2-z_j^2}$, i.e. the section of the bead at $z_j$. Note that the way the sum is implemented, using $N=1$ (i.e. one slice through the center of the bead) corresponds to a convolution of the PSF with a disk of the same nominal radius as the bead, giving a very simple model for the blurring effect.

Alternatively, a semi-analytic approach can be used based on a Taylor expansion in $z$ around $z_0$ of $I_{\mathrm{det}}$ like that in Eq.~(\ref{eq:TaylorSeries}), 
namely, 
\begin{equation}
	I_{\mathrm{tot}}(\boldsymbol{\bar{\rho}};x_0,y_0,z_0,R)
	=  \sigma_V \sum_{m=0}^{\infty} \frac{1}{m!}\mathcal{I}_{\mathrm{det},m}(\boldsymbol{\bar{\rho}};x_0,y_0,z_0) \ast G_m(\boldsymbol{\bar{\rho}};R),
	\label{eq:full_sphere_semianalytic}
\end{equation}
where, for sufficiently small beads, a good approximation can be obtained by truncating the sum at a low value of $m$. The function $G_m(\boldsymbol{\bar{\rho}};R)$ is defined as the easily-solvable integral
\begin{equation}\label{eq:Gdefinition}
	G_m(\boldsymbol{\bar{\rho}};R) = \int^{R}_{-R} z^m D(\boldsymbol{\bar{\rho}};R,z) dz
	= \int^{\ell/2}_{-\ell/2} z^m dz
	=\left\{\begin{array}{cc}\frac{1}{2^m(m+1)}\ell^{m+1},&m\,{\rm even,}\\
		0,&m\,{\rm odd}.\end{array}\right.
\end{equation}
where $\ell(\boldsymbol{\bar{\rho}};R)$ is the length in $z$ of the bead at a given transverse position. In terms of the detector coordinates, this length is given by
\begin{equation}\label{eq:def_ell}
	\ell({\bar{\rho}};R)=2\sqrt{R^2-\left|\frac{{\bar{\rho}}}{M}\right|^2}.
\end{equation}
Note that $G_m=0$ for odd $m$ due to the symmetry of the bead, so the sum in Eq.~(\ref{eq:full_sphere_semianalytic}) includes only contributions for even $m$, even when $n_0\neq n_f$. We therefore substitute $m=2\mu$ for $\mu=0,1,...$ in what follows. 
By using the convolution property of Fourier transforms, we can rewrite Eq.~(\ref{eq:full_sphere_semianalytic}) as
\begin{equation}
	I_{\mathrm{tot}}= \sigma_V~\mathcal{F}^{-1}_{\mathbf{u}\rightarrow\boldsymbol{\bar{\rho}}}\left\{ \sum_{\mu=0}^\infty\frac{1}{(2\mu)!} \mathcal{F}_{\boldsymbol{\bar{\rho}}\rightarrow\mathbf{u}}\left\{\mathcal{I}_{\mathrm{det},2\mu}\right\} \mathcal{F}_{\boldsymbol{\bar{\rho}}\rightarrow\mathbf{u}}\{G_{2\mu}\} \right\} 
	\label{eq:semianalytic}
\end{equation}
where we dropped the arguments for brevity. Since $G_{2\mu}$ are radially symmetric functions, the Hankel transform can be used to obtain their Fourier transforms:
\begin{equation}\label{eq:Fourier_G}
	\mathcal{F}_{\boldsymbol{\bar{\rho}}\rightarrow\mathbf{u}}\{G_{2\mu}\}(u) = 4\pi~ M^{2\mu+2}R^{2\mu+3}(2\mu-1)!!~\frac{j_{\mu+1}(ku R)}{(ku R)^{\mu+1}},
\end{equation}
where $n!!=(n)(n-2)\dots$ is the double factorial, and $j_n$ denotes a spherical Bessel function of order $n$. 

\begin{figure}[ht]
	\centering
	\includegraphics[scale=0.87]{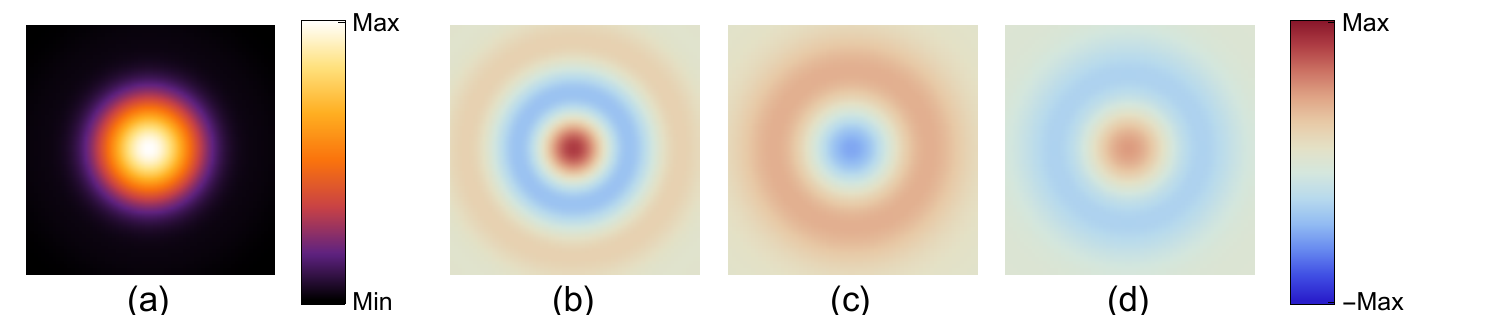}
	\caption{(a) Simulated PSFs for a 200nm diameter fluorescent bead using numerical integration with 21 slices.  Difference between the numerical integration result and (b) one slice, i.e. uniform disk model, (c) semi-analytic model with $\mu=0$, and (d) semi-analytic model with $\mu=1$. Note that (b) and (c) are normalized to $1\%$ of the peak value of the PSF, whereas (d) is normalized to $0.05\%$ of the maximum value recorded for (a). Here $\mathrm{NA}=1.3$, and the index-matching condition  ($n_f=n_0\approx1.33$) and squares are of the length of $1\mu$m on the object side.}
	\label{fig:8}   
\end{figure}

Figure \ref{fig:8}(a) shows the PSF for the particular case of a bead with diameter $2R=200$nm embedded in water and index matched, imaged by a system with NA$ = 1.3$. The PSF was obtained using the discrete sum of convolutions in Eq.~(\ref{eq:sum_disks}) with $N=21$. On the other hand, Fig. \ref{fig:8}(b-c) shows the difference between our ground truth, the PSF in Fig. \ref{fig:8}(a) and the PSFs obtained using (b) the same method but for $N=1$, namely a simple convolution with a uniform disk, and (c,d) the semi-analytical approach in Eq.~(\ref{eq:semianalytic}) with the sum truncated at (c) $\mu=0$, namely a simple convolution with the apodized disk $G_0$, and (d) $\mu=1$, that is, including an extra correction. We take as ground truth the result shown in Fig. \ref{fig:8}(a). We see that, for large NAs the two models including only one contribution, namely the convolution with a uniform disk $D(\boldsymbol{\bar{\rho}})$ (b) and with an apodized disk $G_0(\boldsymbol{\bar{\rho}})$ (c), give results with similar levels of error,
whereas the semi-analytic result including a correction term gives an error 20 times smaller. Note that the errors in Fig. \ref{fig:8}(b) and (c) have opposite behavior, due to the fact that $G_0(\boldsymbol{\bar{\rho}})$ weights more heavily the center compared to the borders, while $D(\boldsymbol{\bar{\rho}})$ is uniform.

\begin{figure}[ht]
	\centering
	\includegraphics[scale=0.79]{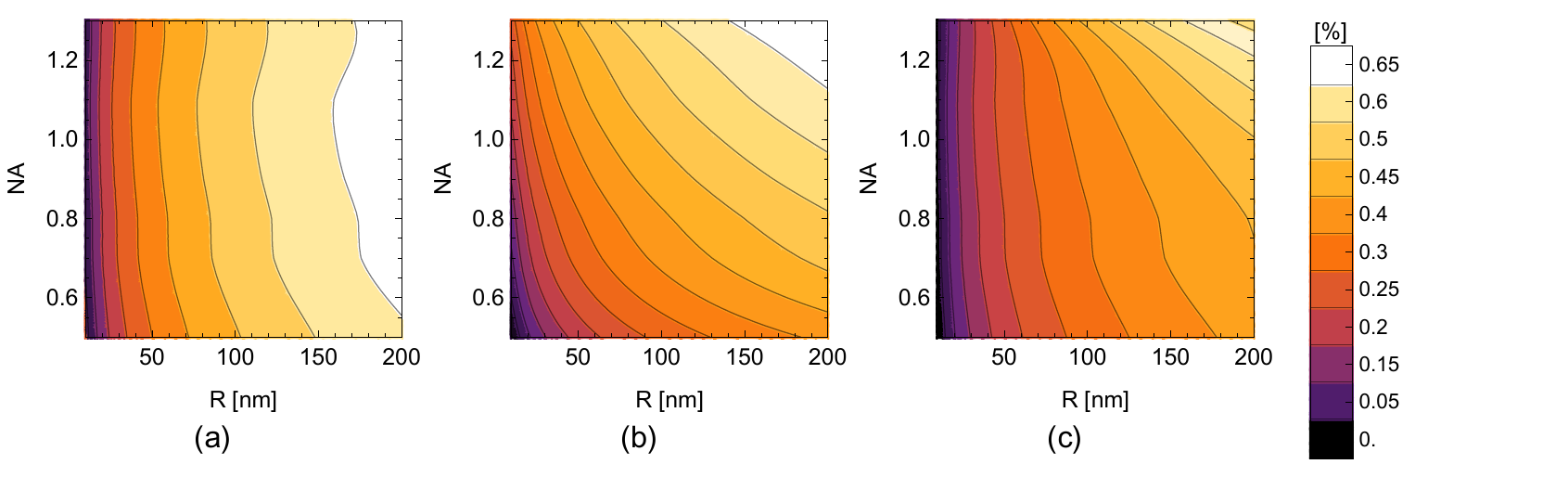}
	\caption{RMS error, for varying NA and bead radius, between the numerically integrated PSF and the PSFs obtained by using: (a) Eq.~(\ref{eq:sum_disks}) with $N=1$(convolution with a flat disk), (b) Eq.~(\ref{eq:semianalytic}) truncated at $\mu=0$ (convolution with an apodized disk), and (c) Eq.~(\ref{eq:semianalytic}) truncated at $\mu=1$. Here $\mathrm{NA}=1.3$ and the embedding medium is water ($n_0\approx1.33$). 
	}
	\label{fig:9}   
\end{figure}

In order to compare the models more thoroughly, the RMS error for a range of NAs and bead radii was computed using again as ground truth the results of Eq.~(\ref{eq:sum_disks}) with $N=15$. Figure \ref{fig:9} shows the logarithm of the RMS error. Notice that for small NA the convolution with the apodized disk, $G_0(\boldsymbol{\bar{\rho}})$, is much better than the convolution with a uniform disk, $D(\boldsymbol{\bar{\rho}})$, but for larger NA they have similar performance, and the uniform disk even outperforms the apodized one. An interesting difference between these two simple models is that convolving with $G_0(\boldsymbol{\bar{\rho}})$ leads to an estimate whose error grows with both NA and $R$, while convolving with $D(\boldsymbol{\bar{\rho}})$ leads to an error that is roughly independent of NA. The semi-analytical model becomes again roughly independent of NA for small beads when the leading correction is included.

A key result, particularly for small NA systems, is that the factor for the zeroth-order, namely
\begin{equation}
	\mathcal{F}_{\boldsymbol{\bar{\rho}}\rightarrow\mathbf{u}}\{G_0\}(u)= 4\pi ~M^2R^3~\frac{j_1(ku R)}{(kuR)},
\end{equation}
can be used to directly deconvolve the measured intensity at the detector to obtain a more realistic PSF, i.e.
\begin{equation}
	I'=\mathcal{F}^{-1}_{\mathbf{u}\rightarrow\boldsymbol{\bar{\rho}}}\left\{\frac{\mathcal{F}_{\boldsymbol{\bar{\rho}}\rightarrow\mathbf{u}}\{I_\mathrm{measured}\}}{\mathcal{F}_{\boldsymbol{\bar{\rho}}\rightarrow\mathbf{u}}\{G_0\}} \right\}.
\end{equation}
This deconvolution is reliable under appropriate signal-to-noise ratio conditions, and requires that the zeros of $\mathcal{F}_{\boldsymbol{\bar{\rho}}\rightarrow\mathbf{u}}\{G_0\}$ lie well outside the incoherent pupil radius of 2NA. The first zero of $\mathcal{F}_{\boldsymbol{\bar{\rho}}\rightarrow\mathbf{u}}\{G_0\}$ is at $u'\approx1.43\pi/(kR)$, giving the safety condition for the bead's diameter $2R<0.715\lambda/\mathrm{NA}$. 
In a similar manner, the uniform disk model obtained form considering a single slice in Eq.~(\ref{eq:slices}) provides an alternative to perform the deconvolution. Here, the location of the first zero of $\mathcal{F}_{\boldsymbol{\bar{\rho}}\rightarrow\mathbf{u}}\{D\}$ restricts the parameters to satisfy the slightly more restrictive condition $2R<0.61\lambda/\mathrm{NA}$.

\subsection{Transparent shell}
Although we do not expect this model to be a faithful representation of a fluorescent bead, we include it given its simplicity, and in case it can be useful in a different application. 
Following the same procedure as in the previous section, let us now assume a fluorescent empty shell of radius $R$ centered at $(x_0,y_0,z_0)$. Again, the measured intensity can be calculated as a superposition of 2D convolutions:
\begin{equation}
	I_{\mathrm{tot}}(\boldsymbol{\bar{\rho}};x_0,y_0,z_0,R)=\sigma_S\int^{R}_{-R} I_{\mathrm{det}}(\boldsymbol{\bar{\rho}};x_0,y_0,z_0+z)\ast A(\boldsymbol{\bar{\rho}};R,z) \,dz,
	\label{eq:shell}
\end{equation}
where $\sigma_S$ is the surface density of fluorophores, $A(\boldsymbol{\bar{\rho}};R,z)$ is an impulse ring function representing the magnified perimeter of the sphere at $z$, i.e. 
\begin{equation}
	A(\boldsymbol{\bar{\rho}};R,z)=\delta \left( \sqrt{\frac{\bar \rho^2}{M^2}+z^2}-R \right)=\frac{\delta(z-\ell/2)+\delta(z+\ell/2)}{|z|/R},
\end{equation}
where $\ell$ is defined in Eq.~(\ref{eq:def_ell}). Numerically, this integral in $z$ can be computed through a discrete sum of $N$ slices as
\begin{equation}
	I_{\mathrm{tot}}=\sigma_S~\mathcal{F}^{-1}_{\mathbf{u}\rightarrow\boldsymbol{\bar{\rho}}}\left\{\sum_{j=1}^{N}\mathcal{F}_{\boldsymbol{\bar{\rho}}\rightarrow\mathbf{u}}\left\{ I_{\mathrm{det}}(\boldsymbol{\bar{\rho}};x_0,y_0,z_0+z_j)\right\}\mathcal{F}_{\boldsymbol{\bar{\rho}}\rightarrow\mathbf{u}}\left\{ A(\boldsymbol{\bar{\rho}};R,z_j) \right\} \right\}\Delta z,
\end{equation}
where $\Delta z$ and $z_j$ are defined as in Eq. (\ref{eq:deltaz}) and
\begin{equation}
	\mathcal{F}_{\boldsymbol{\bar{\rho}}\rightarrow\mathbf{u}}\left\{ A(\boldsymbol{\bar{\rho}};R,z_j) \right\}(u)=\mathcal{F}_{\boldsymbol{\bar{\rho}}\rightarrow\mathbf{u}}\left\{ \frac{M R}{\bar \rho}\delta\left(\frac{\bar \rho}{M}- R_j \right)\right\}(u)=2\pi M^2 R ~J_0(kuR_j),
\end{equation}
in which once more $R_j=\sqrt{R^2-z_j^2}$, i.e. it is the radius of the bead at $z_j$.

On the other hand, as in the previous section, a semi-analytic approach can be used based on a Taylor expansion in $z$ around $z_0$ of $I_{\mathrm{det}}$, namely, 
\begin{equation}
	I_{\mathrm{tot}}(\boldsymbol{\bar{\rho}};x_0,y_0,z_0,R)
	=  \sigma_S\sum_{m=0}^{\infty} \frac{1}{m!}\mathcal{I}_{\mathrm{det},m}(\boldsymbol{\bar{\rho}};x_0,y_0,z_0) \ast H_m(\boldsymbol{\bar{\rho}};R),
	\label{eq:shell_semianalytic}
\end{equation}
where the function $H_m(\boldsymbol{\bar{\rho}};R)$ is defined as the easily-solvable integral
\begin{equation}
	H_m(\boldsymbol{\bar{\rho}};R) = \int^{R}_{-R} z^m A(\boldsymbol{\bar{\rho}};R,z) dz
	= \left\{\begin{array}{cc}R~\frac{\ell^{m-1}}{2^{m-2}},&m\,{\rm even,}\\
		0,&m\,{\rm odd}. \end{array}\right.
\end{equation}
Note that in general $H_m=(m-1)RG_{m-2}$, where $G_m$ is given in Eq.~(\ref{eq:Gdefinition}), and so the total intensity can be expressed (by again substituting $m=2\mu$) as
\begin{equation}
	I_{\mathrm{tot}}=R \sigma_S~\mathcal{F}^{-1}_{\mathbf{u}\rightarrow\boldsymbol{\bar{\rho}}}\left\{ \sum_{\mu=0}^\infty\frac{(2\mu-1)}{(2\mu)!} \mathcal{F}_{\boldsymbol{\bar{\rho}}\rightarrow\mathbf{u}}\left\{\mathcal{I}_{\mathrm{det},2\mu}\right\} \mathcal{F}_{\boldsymbol{\bar{\rho}}\rightarrow\mathbf{u}}\{G_{2(\mu-1)}\} \right\}, 
\end{equation}
in which $\mathcal{F}_{\boldsymbol{\bar{\rho}}\rightarrow\mathbf{u}}\{G_{2\mu}\}$ is given in Eq. (\ref{eq:Fourier_G}).

\section{Mimicking transverse dipoles with beads}
\label{sec:mimicking}
We now evaluate how well a flourescent bead supplemented with a polarizer at the pupil can be used to mimic the PSF of a transverse point dipole. 
For of a transverse dipole, the intensity distribution at the pupil is fairly uniform except near the pupil edge and is similar to an unpolarized emitter.
Therefore, by inserting a linear polarizer at the pupil, one can make the PSF of an unpolarized emitter mimic the one for a transverse dipole \cite{zhang:2016, Curcio:2020}. For systems with small NA, this should be a good approximation, while for higher NAs differences become more evident at the edge of the pupil distribution, both in intensity and polarization, causing discrepancies between the corresponding PSFs. In this section we first study how a point-like emitter can mimic a real transversal dipole, and then we incorporate the effects of bead size using the results of the previous section. Particularly we highlight how the function $G_0$ can be used to deconvolve the measured PSFs to correct the blurring introduced by the bead size.

\subsection{Mimicking transverse dipoles using a point-like source}
Let us, without loss of generality, consider 
a horizontal dipole (aligned with the $x$ axis). The Jones matrix at the pupil is then written as
\begin{equation}
	\mathbb{J}(\mathbf{u})=\mathbb{J}'(\mathbf{u})\mathbb{P}_x,
\end{equation}
where $\mathbb{P}_x=\boldsymbol{\hat{\rho}_x}\boldsymbol{\hat{\rho}_x}^\dagger$ is a horizontal projection matrix representing a horizontal polarizer, with $\boldsymbol{\hat{\rho}_x}$ being a 2D unit vector in the horizontal direction. That is, the Jones matrix representing all the manipulations after the pupil can be chosen to represent the cascade of a horizontal linear polarizer 
and other possible manipulations represented by the Jones matrix $\mathbb{J}'(\mathbf{u})$. Let us for now set this second Jones matrix to the identity. It can be shown that the intensity distribution at the pupil just after the polarizer becomes
\begin{equation}
	\begin{aligned}
		I_\mathrm{p,mimic}(\mathbf{u})=&\left(|g_0(u)|^2+|g_2(u)|^2+2~\mathrm{Re}\,\{g_0^*(u)g_2(u)\}\cos2\varphi\right.\\
		&\left. +|g_1(u)|^2\cos^2\varphi\right)F(u;z_0),
	\end{aligned}
\end{equation}
whose only difference with that for a true horizontal dipole is the last term, $|g_1(u)|^2\cos^2\varphi$, meaning that the difference is zero only at the central axial line perpendicular to the polarization.

\begin{figure}[ht]
	\centering
	\includegraphics[scale=0.9]{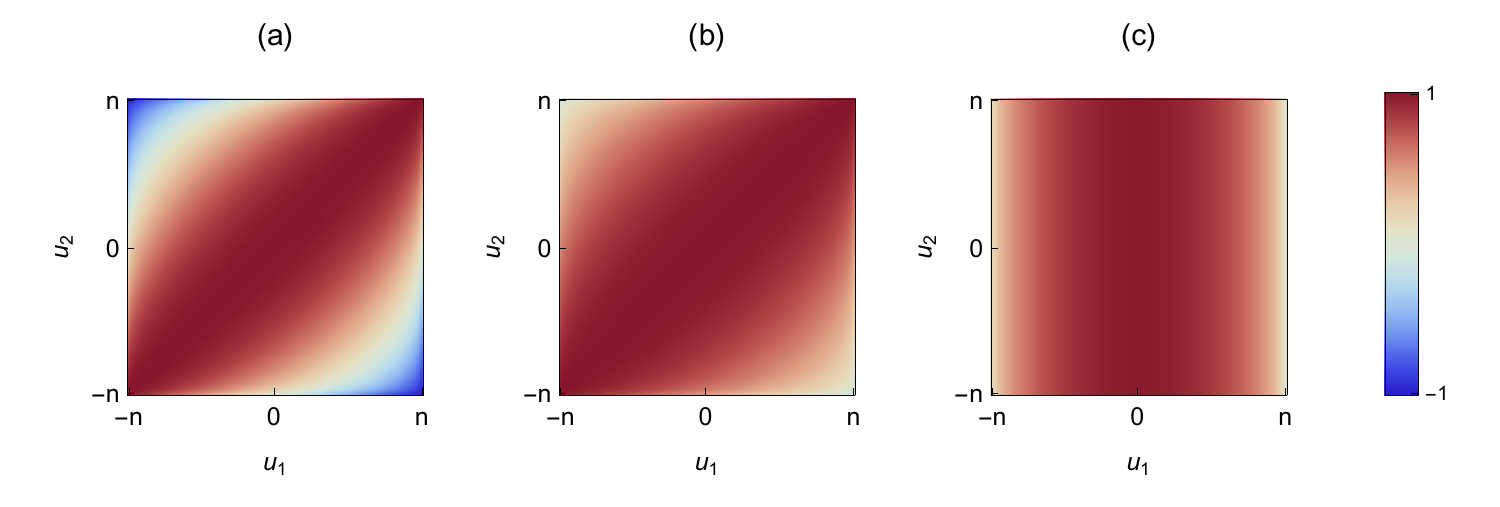}
	\caption{Degree of coherence $\mu_{\mathrm{p,mimic}}$ for (a) $\varphi_1=\varphi_2=0$, (b) $\varphi_1=\varphi_2=\pi/4$ and (c) $\varphi_1=0$ and $\varphi_2=\pi/2$. Here the unpolarized emitter is assumed to be in the water ($n_0\approx1.33$), $n_f=1.51$, and we only show $u_{1,2}$ within the non-SAF regions, since for SAF the degree of coherence becomes complex.}
	\label{fig:10}   
\end{figure}

Perhaps of more consequence to the final PSFs are the differences in spatial coherence and polarization over the pupil. For a pure horizontal dipole source, the cross-spectral density matrix is
\begin{equation}
	\mathbb{W}_x({\bf u}_1,{\bf u}_2)=\mathbb{K}({\bf u}_1)~\mathbf{\hat{x}}\mathbf{\hat{x}}^\dagger~\mathbb{K}^\dagger({\bf u}_2),
\end{equation}
where $\mathbf{\hat{x}}\mathbf{\hat{x}}^\dagger$ is a 3D projection in the $x$ direction (as opposed to $\mathbb{P}_x$ which enacts a 2D projection). Note that $\mathbb{W}_x$ is simply the outer product of  $\mathbb{K}~\mathbf{\hat{x}}$ and its conjugate (each evaluated at a different pupil point), so there is no statistical decorrelation between any two pupil points.
This is not the case, on the other hand, for the mimicked dipole based on an unpolarized emitter followed by a polarizer at the pupil, for which all three orthogonal dipole directions contribute to the $x$ component at the pupil, each with a different distribution (two of them being important mainly at the edges of the pupil). These contributions are mutually incoherent. Therefore, the cross-spectral density matrix at the pupil right after the polarizer due to an unpolarized emitter at the nominal source plane is given by
\begin{equation}\label{eq:cross-spectral}
	\mathbb{W}_\mathrm{p,mimic}({\bf u}_1,{\bf u}_2)=\mathbb{P}_x\mathbb{K}({\bf u}_1) \mathbb{K}^\dagger({\bf u}_2)\mathbb{P}_x
\end{equation}
whose only nonzero element is the $xx$ one. The degree of coherence for this nonzero component is then
\begin{equation}
	\mu_\mathrm{p,mimic}({\bf u}_1,{\bf u}_2)=\frac{\{\mathbb{W}_\mathrm{p,mimic}({\bf u}_1,{\bf u}_2)\}_{xx}}{\sqrt{\{\mathbb{W}_\mathrm{p,mimic}({\bf u}_1,{\bf u}_1)\}_{xx}\{\mathbb{W}_\mathrm{p,mimic}({\bf u}_2,{\bf u}_2)\}_{xx}}}.
\end{equation}
Note that for an emitter at the origin and in the absence of aberrations or SAF contributions, this degree of polarization is real. Because $\{\mathbb{W}_\mathrm{p,mimic}\}_{xx}$ is not a separable function of $\mathbf{u}_1$ and $\mathbf{u}_2$, $\mu_\mathrm{p,mimic}$ typically differs from unity for $\mathbf{u}_1\ne\mathbf{u}_2$. This is illustrated in Fig. \ref{fig:10} for different slices of the pupil: two in which $\mathbf{u}_1\parallel\mathbf{u}_2$, and one in which $\mathbf{u}_1\perp\mathbf{u}_2$. Note in particular that for sufficiently separated points along the horizontal line, the degree of coherence vanishes and then becomes negative, due to the predominance at those points of the radially polarized contribution introduced by the longitudinal dipole. On the other hand, for any two pupil points along a vertical line ($\varphi_1=\varphi_2=\pi/2$), the polarizer eliminates the contributions from the vertical and longitudinal components, so that $\mu_\mathrm{p,mimic}=1$.  In general, if both points are near the center of the pupil, the correlation approaches unity.

A quantitative measure of the similarity of both pupil distributions that accounts for amplitude, coherence and polarization is given by the following normalized projection: 
\begin{align}
	C({\rm NA})&=\left\{\frac{\iint{\rm Tr}[\mathbb{W}_x({\bf u}_1,{\bf u}_2)\mathbb{W}_\mathrm{p,mimic}({\bf u}_2,{\bf u}_1)]\,d^2{\bf u}_1\,d^2{\bf u}_2}{\int{\rm Tr}[\mathbb{W}_x({\bf u},{\bf u})]\,d^2{\bf u}\,\int{\rm Tr}[\mathbb{W}_\mathrm{p,mimic}({\bf u},{\bf u})]\,d^2{\bf u}}\right\}^{1/2}. 
\end{align}
This measure achieves its maximum value of unity only if the two cross-spectral density matrices are identical. 
Given Parseval's identity, this measure also quantifies the difference in amplitude, coherence and polarization of the corresponding fields at the detector plane; given appropriate normalization, such RMS difference is given by $\sqrt{1-C^2}$. However, the main interest in this work is in differences in PSF, namely only in intensity. Therefore, it is expected that $\sqrt{1-C^2}$ will overestimate (and hence provide an upper bound for) the RMS errors in PSF.
It is convenient to use polar coordinates to evaluate this measure, and for the case considered here the integrals in the angular variables turn out to be solvable in closed form, leading to the expression
\begin{align}
	C({\rm NA})
	&=\frac{\mathcal{G}_0+\mathcal{G}_2/2}{\sqrt{(\mathcal{G}_0+\mathcal{G}_2)(\mathcal{G}_0+\mathcal{G}_1/2+\mathcal{G}_2)}},
\end{align}
where $\mathcal{G}_j=\int |g_j(u)|^2F(u,z_0)\,u\,du$. Note that the dependence in $z_0$ remains for the SAF region. 


So far we have considered the similarity of the PSFs for a transverse dipole and an unpolarized source (supplemented with a polarizer), both at the focal plane. Let us consider now the effect of a small amount of defocus.
The Taylor expansion approach is useful for 
these purposes. 
Figure \ref{fig:7}(b) shows the distributions of $\mathcal{I}_m$ at the nominal focal distance for an unpolarized emitter followed by a horizontal polarizer. Given their uniform polarization, these PSF components maintain their distributions (except for a scaling factor) when projected onto any other polarization. Therefore, unlike a single horizontal dipole, as shown in Fig.~\ref{fig:4}(d,e), $\mathcal{I}_m$ remains zero for odd $m$ even when projecting onto a circular polarization, showing that the behavior of a transverse dipole under defocus is not being fully mimicked.

\subsection{Mimicking transverse dipoles using finite-size beads}

\begin{figure}[ht]
	\centering
	\includegraphics[scale=0.7]{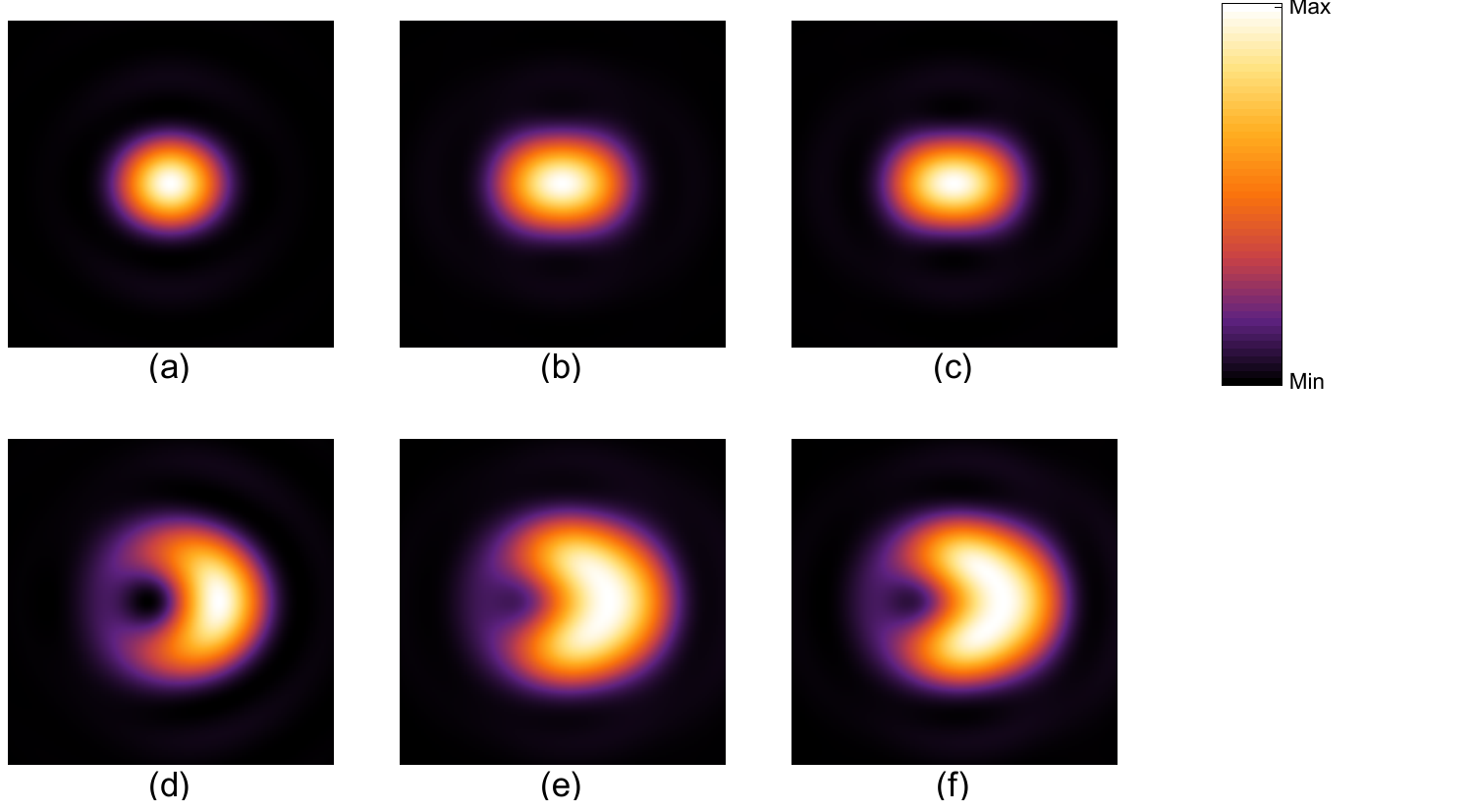}
	\caption{Simulation of the PSFs of (a,d) true horizontal dipoles, (b,e) mimicked ones using a 200nm diameter fluorescent bead and a polarizer, and (c,f) the deconvolution of the results in (b) and (e) with an apodized disk $G_0$. These PSFs are (a-c) for no polarization filtering, and (d-f) for an engineered PSF \cite{Curcio:2020}. Here NA$=1.3$, there is index matching and the embedding medium is water ($n_f=n_0\approx1.33$). Squares represent $1.5\mu$m in the object space.
	}
	\label{fig:11}   
\end{figure}

Let us finish by comparing the PSF of a true horizontal dipole to that of a mimicked one using a fluorescent bead and a linear polarizer at the pupil plane of the imaging system. For the latter, we use the approximation in Eq.~(\ref{eq:sum_disks}) with $N=15$. We also test the deconvolution method discussed in the previous section to compensate for the blurring introduced by the bead's size. Figure \ref{fig:11} shows the resulting simulated PSFs for a high NA system with no SAF and with an index matching scenario (NA$<n_0=n_f$) for two cases: a standard imaging system with no aberrations, and a PSF engineering technique called CHIDO \cite{Curcio:2020} based on a spatially-varying birefringent mask and circular polarization filtering.

\begin{figure}[ht]
	\centering
	\includegraphics[scale=0.8]{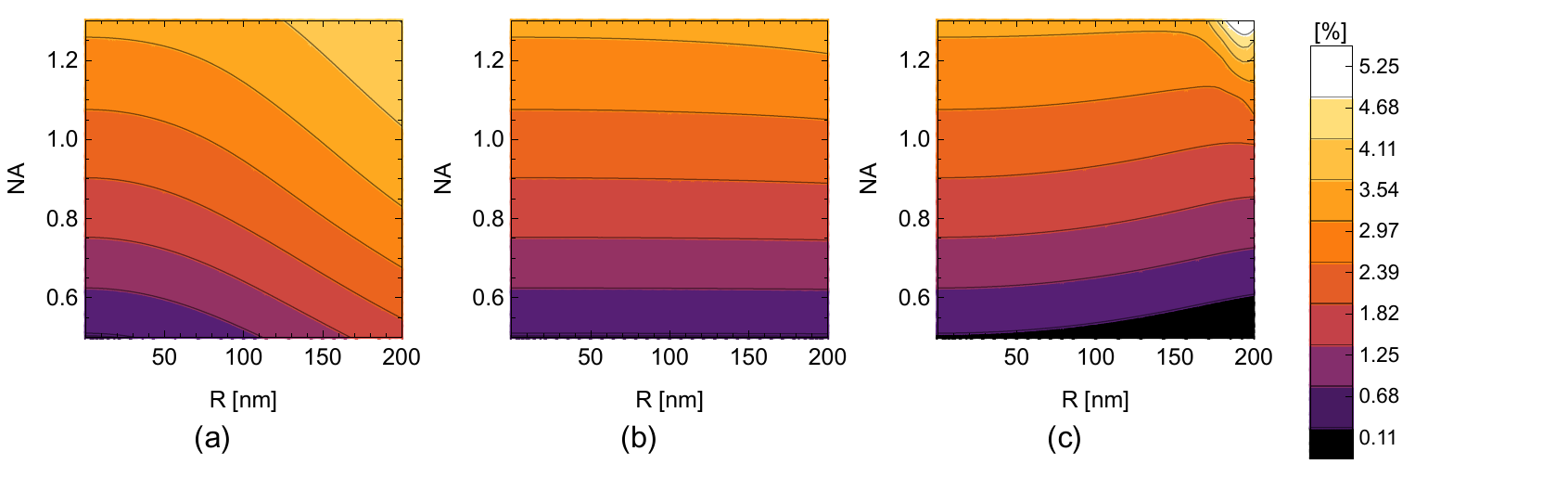}
	\caption{RMS error, for varying NA and bead radius, between the PSF of a true transverse dipole and the mimicked one using a fluorescent bead for: (a) no treatment, (b) deconvolution with the apodized disc $G_0$, and (c) deconvolution with a uniform disk. Here $\mathrm{NA}=1.3$ and the embedding medium is water ($n_0=1.33$).
	}
	\label{fig:12}   
\end{figure}

We also computed the RMS error between the PSFs of an emulated fluorophore (with and without deconvolution) and a true horizontal dipole using different bead radii and various values of NA; these are shown in Fig.~\ref{fig:12}. 
Naturally, the error between the raw PSFs increases with both NA and bead radius, as shown in Fig.~\ref{fig:12}(a). 
Note from Fig.~\ref{fig:12}(b), though, that the deconvolution using the apodized disk $G_0$ eliminates almost completely the error's dependence on the radius (since the contours are essentially horizontal). Surprisingly, on the other hand, the deconvolution with a uniform disk, shown in Fig.~\ref{fig:12}(c) also accounts well for the size mismatch, and can give an error that is smaller for larger beads. 
We performed the same computations for CHIDO's engineered PSFs and obtained almost identical behavior, only with slightly higher RMS errors (approximately shifted up by $1\%$).

The reason why the effect of particle size is not necessarily negative can be understood from the coherence analysis. Due to the van Cittert-Zernike theorem, the effect of the spatial extension of the bead is to include an extra factor in the cross-spectral density matrix at the pupil given in Eq.~(\ref{eq:cross-spectral}). This factor turns out to take a simple analytic form for the two bead models discussed here: for the solid sphere it is given by  $3\sigma_V\,j_1(\eta)/\eta$, while for the empty shell it is given by $\sigma_S{\rm sinc}(\eta)$, where for an index matching scenario
\begin{equation}
	\eta({\bf u}_1,{\bf u}_2)=knR\sqrt{u_1^2+u_2^2-2u_1u_2\cos(\varphi_2-\varphi_1)+[\gamma^*(u_1,n)-\gamma(u_2,n)]^2}. 
\end{equation} 
Note that for $R\to0$ the two factors tend to a constant equal to $\sigma_V$ or $\sigma_S$. For $R>0$, on the other hand, these factors decrease as $u_1$ and $u_2$ increase, meaning that the contributions from the edges of the pupil are partially suppressed. These are precisely the regions where the radiation pattern of a transverse dipole differs most from that of an unpolarized source supplemented with a linear polarizer, and therefore this suppression is not necessarily detrimental to the fit. As shown by the results shown in Fig.~\ref{fig:12}(c), increasing the radius can even lead to possibly better matches, particularly after post-processing of the PSFs. 

\section{Conclusions}
\label{sec:conclusions}


In this work we studied how and under which conditions fluorescent beads can be used to mimic transverse dipoles. We did this by developing models that take into account the three-dimensional size and the incoherent and isotropic emission pattern of fluorescent beads. Of particular importance is the development of a semi-analytic method to model the blurring due the bead size. This method is based on a Taylor expansion around the bead's central plane and provides different orders of corrections for the final PSF which are implemented through two-dimensional convolutions. It is therefore much easier to implement than a full three-dimensional convolution, and it converges rapidly, with the first two terms providing already an estimate of the blurring that is almost indistinguishable from the exact result for typical bead sizes and NAs.  
Additionally, we showed that the difference in size between a point-like incoherent source and a fluorescent bead can be corrected satisfactorily by 2D deconvolution with the apodized disk function given in Eq. (\ref{eq:Gdefinition}), hence reducing drastically the computational resources needed for post-processing. 

The spatial extent of the bead is not the only factor that makes its PSF differ from that of a point dipole. The second important difference is in their radiation patterns. The pattern for the bead can be made more similar to that of the dipole by placing at the pupil a polarizer plane that is aligned with the direction of the dipole, assumed here to transverse to the axis of the microscope. For systems with small NA this leads to very good agreement, but for larger NA there are notable differences in the intensity, polarization direction, and even the spatial coherence of the pupil distributions of the point dipole and the bead followed by a polarizer. These differences result in a difference in the corresponding PSFs, particularly in their high spatial frequency components. As mentioned earlier, the difference due to the blurring caused by bead size can be partially removed through deconvolution. 
Surprisingly, the resulting error can sometimes be smaller for larger beads. This can be explained by the fact that the spatial blurring suppresses the contributions at the pupil plane from the edges of the pupil (i.e., the PSF's high spatial frequencies). The deconvolution process, which aims to reinstate these suppressed contributions, does it in a way that can be sometimes more consistent with the pattern for a point dipole than the suppressed contributions were.  

The current manuscript focuses mostly on transverse dipoles, but future work will extend this study to ways of emulating longitudinal and/or oblique dipoles by using beads and appropriate polarization plates.

	\section*{Funding}
	This research received funding from the 3DPolariSR and 3Dpol ANR grants (ANR-20-CE42-0003, ANR-21-CE24-0014). R.~G.~C. acknowledges funding from the Labex WIFI (ANR-10-LABX-24, ANR-10-IDEX-0001-02 PSL*).

	
	\section*{Disclosures}
	The authors declare no conflicts of interest.
	
	\section*{Data Availability Statement}
	The code used is available from the corresponding authors upon reasonable request.
	

\bibliographystyle{ieeetr}
\bibliography{sample}

\end{document}